# Challenges in designing appropriate scaffolding to improve students' representational consistency: The case of a Gauss's law problem

Alexandru Maries,[1] Shih-Yin Lin,[2] and Chandralekha Singh[3]

[1]*Department of Physics, University of Cincinnati, Cincinnati, Ohio 45221, USA*
[2]*Department of Physics, National Changhua University of Education, Changhua 500, Taiwan*
[3]*Department of Physics and Astronomy, University of Pittsburgh, Pittsburgh, Pennsylvania 15260, USA*

Prior research suggests that introductory physics students have difficulty with graphing and interpreting graphs. Here, we discuss an investigation of student difficulties in translating between mathematical and graphical representations for a problem in electrostatics and the effect of increasing levels of scaffolding on students' representational consistency. Students in calculus-based introductory physics were given a typical problem that can be solved using Gauss's law involving a spherically symmetric charge distribution in which they were asked to write a mathematical expression for the electric field in various regions and then plot the electric field. In study 1, we found that students had great difficulty in plotting the electric field as a function of the distance from the center of the sphere consistent with the mathematical expressions in various regions, and interviews with students suggested possible reasons which may account for this difficulty. Therefore, in study 2, we designed two scaffolding interventions with levels of support which built on each other (i.e., the second scaffolding level built on the first) in order to help students plot their expressions consistently and compared the performance of students provided with scaffolding with a comparison group which was not given any scaffolding support. Analysis of student performance with different levels of scaffolding reveals that scaffolding from an expert perspective beyond a certain level may sometimes hinder student performance and students may not even discern the relevance of the additional support. We provide possible interpretations for these findings based on in-depth, think-aloud student interviews.

## I. INTRODUCTION

Physics is a challenging subject to learn and it is difficult for introductory students to associate the abstract concepts they study in physics with more concrete representations that facilitate understanding without an explicit instructional strategy aimed to aid them in this regard. Without guidance, introductory students often employ formula oriented problem-solving strategies instead of developing a solid grasp of physical principles and concepts [1]. There are many reasons to hypothesize that multiple representations of concepts along with the ability to construct, interpret, and transform between different representations that correspond to the same physical system or process play a positive role in learning physics. First, physics experts often use multiple representations as a first step in a problem-solving process [1–4], and diagrammatic representations have been shown to be superior to exclusively employing verbal representations when solving problems [5–7]. Second, students who are taught explicit problem-solving strategies emphasizing the use of different representations of knowledge at various stages of problem solving construct higher-quality and more complete representations and perform better than students who learn traditional problem-solving strategies [8]. Third, multiple representations are very useful in translating the initial, mostly verbal description of a problem into a representation more suitable to mathematical manipulation [9,10], partly because the process of constructing a representation of a problem makes it easier to generate appropriate decisions about the solution process. Also, getting students to represent a problem in different ways helps shift their focus from merely manipulating equations toward understanding physics [11]. Some researchers have argued that in order to understand a physical concept thoroughly, one must be able to recognize and manipulate the concept in a variety of representations [10,12]. As Meltzer puts it [12], a range of diverse representations is required to "span" the conceptual space associated with an idea. Since traditional courses that do not emphasize multiple representations lead to low gains on the Force Concept Inventory [13,14] and on other assessments in the domain of electricity and magnetism [15], in order to improve students' understanding of physics concepts, many researchers have developed instructional strategies that place explicit emphasis on multiple representations [1,9,16–18] while other researchers have developed other strategies with implicit focus on multiple representations [19–21]. Van Heuvelen's approach [9], for example, starts by ensuring that students explore the qualitative nature of concepts by using a variety of representations of a concept in a familiar setting before adding the complexities of mathematics. Many other researchers have emphasized the importance of students becoming facile in translating between different representations of knowledge [22,23] and that significant positive learning occurs when students develop facility in the use of multiple forms of representation [24]. However, careful attention must be paid to instructional use of diverse representational modes since specific learning difficulties may arise as a consequence [12] because students can approach the same problem posed in different representations differently without support [12,25].

This paper is focused on students' ability to transform between mathematical and graphical representations of a piecewise function in different regions in the context of an electrostatics problem with spherically symmetric charge distribution and the effect of different scaffolding supports designed to aid them in this regard. Student difficulties in interpreting graphical representations have been extensively researched in kinematics [26–30]. Instructional strategies have also been developed to reduce student difficulties [31–33]. Other researchers have investigated student understanding of $P$-$V$ (pressure versus volume) diagrams both in upper-level thermodynamics courses [34] as well as in introductory calculus-based physics courses [35]. In a later study, Christensen and Thompson [36] investigated student difficulties with the concept of slope and derivative in a mathematical (graphical) context.

Student difficulties with the concept of a function have been researched by mathematics education researchers [37–39]. Hitt [39] found that even secondary mathematics teachers cannot always articulate between the various systems of representation involved in the concept of a function. Vinner and Dreyfus [40] distinguished between a concept image and a concept definition because they saw students repeatedly misuse and misapply terms like function, limit, tangent, and derivative. For many students, the image evoked by the term "function" is of two expressions separated by an equal sign. Thompson found [41] that many students who had successfully passed a calculus and a modern algebra course saw no problem with a definition like $f(x) = n(n + 1)(2n + 1)/6$ because it fits their concept image of a function. Also, students in introductory physics courses often have an action conception of a function because a function is seen as a command to calculate, and therefore they must actually apply it to a number before the "recipe" will produce anything. Without guidance, the way many introductory physics students manage equations in solving physics problems is often very predictable: they plug numbers into an equation and figure out an unknown, which can in turn be plugged into another equation. This "plug and chug" process is continued until the target variable is found. When numbers are not given or when students run into a situation with two equations and two unknowns, they have much more difficulty solving the problem. As evidenced by these examples and others [41], while students' concept images are often not consistent with concept definitions, for mathematics experts, the concept images become tuned over time so that they are consistent with the conventionally accepted concept definitions. One proposed instructional method of overcoming some of these difficulties involves real-world investigations that use realistic data and scenarios [42–44]. Mathematics education researchers have also investigated student difficulties in connecting various representations of functions, in particular, graphical and algebraic representations [45,46]. Some researchers have emphasized that this process of translating between the graphical and algebraic representations of functions presents one of the central difficulties for students in constructing an appropriate mental image of a function [47]. Other mathematics education researchers have investigated the intertwining between the flexibility of moving from one representation of a function to another and other aspects of knowledge and understanding [48–50] as well as students' abilities to extract meaningful information from graphs [51].

In physics, there is the added difficulty of understanding the relevance of certain mathematical knowledge and procedures to the solution of physical problems. Students may have the requisite mathematical knowledge that needs to be applied to a physical situation, but they may fail to invoke it at the appropriate time because they are unaware of its usefulness. This is supported by Hammer's observation that high-school students take little out of an initial mathematical review of procedures divorced from physics [52] and by research on difficulties of transferring mathematical knowledge across disciplines [53–55]. Also, the physics context typically requires additional information processing, which may lead to an increased cognitive load [56] and deteriorated performance.

Here, we explore the facility of students in a calculus-based introductory physics course in transforming a problem solution involving the electric field for spherical charge symmetry from a mathematical to a graphical representation, and the effect of different scaffolding supports on students' ability to carry out the transformation consistently. This study is primarily focused on students' ability to transform electric field from one representation to another and not on their ability to correctly use Gauss's law to find the electric field. However, many previous studies have documented students' difficulties with E&M (electricity and magnetism) concepts [57–71]. In study 1, we investigated the extent to which students were able to transform from one representation to another consistently, and we conducted think-aloud interviews with students to identify common difficulties. In study 2, we designed two scaffolding support levels that built on each other based on the findings of study 1, and investigated their impact on improving students' representational consistency.

## II. STUDY 1

### A. Methodology and research questions

Since being able to transform between different representations of knowledge is a hallmark of expertise, we investigated the extent to which students in a calculus-based introductory physics course could transform the electric field in each region for the situation depicted in Fig. 1 from a mathematical to a graphical representation. We selected this problem which is relatively straightforward from an expert point of view [since the electric field is no-zero only in region (ii)]. Despite its simplicity for experts, our past experience with this problem indicated that students have great difficulty with this problem. In particular, roughly 70% of the students in this study found a nonzero expression for the electric field in at least one region in which it is zero. Moreover, many students (including those who found correct expressions for the electric field in each region) had difficulty in transforming the electric field in each region from a mathematical to a graphical representation.

Study 1 was designed to investigate the extent to which students have difficulty in transforming the piecewise electric field in this problem from a mathematical to a graphical representation, and to identify possible reasons that could account for the common difficulties they exhibit when they engage in the task. In this phase of the investigation, the problem was administered as a quiz to an introductory calculus-based class of 65 students. This was one of the weekly quizzes students took at the end of recitation which was counted for a certain (small) percentage of their final grade. In addition, in order to identify possible difficulties students have in plotting the electric field in this situation, think-aloud interviews [72] were conducted with eight students individually. The interviews suggested possible student difficulties. The quantitative data were then used to estimate the prevalence of each type of difficulty identified. Then, these findings inspired study 2, in which two levels of scaffolding support that built on each other (i.e., the second level builds on the first) were designed to help reduce the most common difficulty observed and guide students to make better connection between the mathematical and the graphical representations of the electric field.

*Problem used in this study.*—"A solid conductor of radius $a$ is inside a solid conducting spherical shell of inner radius $b$ and outer radius $c$. The net charge on the solid conductor is $+Q$ and the net charge on the concentric spherical shell is $-Q$ (see Fig. 1).

(a) Write an expression for the electric field in each region.

 (i) $r < a$
 (ii) $a < r < b$
 (iii) $b < r < c$
 (iv) $r > c$

(b) On the figure below (see Fig. 2), plot $E(r)$ (which is the electric field at a distance $r$ from the center of the sphere) in all regions for the problem in (a).

We investigated the following research questions:

**RQ1: Without any scaffolding support, to what extent are students able to make connections between the mathematical and graphical representations of the electric field in this problem?**—This research question was answered by determining the percentage of students who plotted the electric field in a manner consistent with the mathematical expressions they found in each region. We considered a students' graph to be consistent with their mathematical expression in a particular region if the shape of the graph agreed with the mathematical expression (e.g., linear, inverse); students were not expected to label end points, or even have correct end points. For example, one student found $E(r) = kr/3$ in region $b < r < c$, and drew a plot similar to the one shown in Fig. 3 (an increasing linear plot that starts from the $r$ axis). Based on the expression he

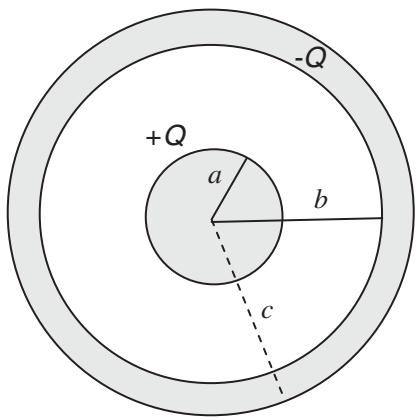

FIG. 1.　Problem diagram.

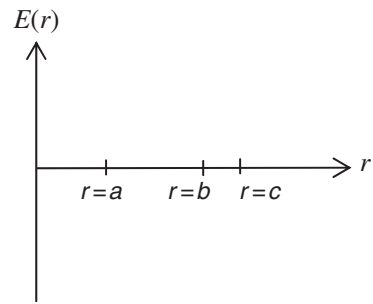

FIG. 2.　Coordinate axes provided to all students for plotting the electric field in part (b).

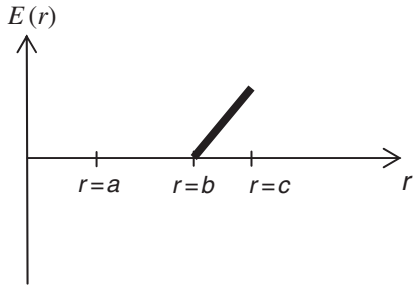

FIG. 3. Reproduction of a plot by a student in region $b < r < c$. His plots in other regions have been excluded for clarity.

wrote down $[E(r) = kr/3]$, the left end point, $E(r = b)$, should be $kb/3$. However, according to his plot in this region (see Fig. 3), the left end point was drawn as $E(r = b) = 0$. Even though the student did not draw the left end point correctly, this student was still considered to be consistent in plotting by the researchers because he drew the correct type of plot (linear) consistent with the expression in that region.

**RQ2: What common difficulties do students have in transforming the electric field in this problem from the mathematical to the graphical representation and how common are these difficulties?**—To answer this research question, we conducted think-aloud interviews during which students solved the problem while verbalizing their thought process. After the students had solved the physics problem to the best of their ability, they were asked for clarification on points they had not made clear earlier while thinking out loud. We identified several difficulties after which we analyzed our quantitative data to determine how common the difficulties are.

Student volunteers were recruited after instructors who were teaching a separate section of the second semester calculus-based introductory physics course sent email announcements to their students with details about the study. The interviews were conducted after students were tested on the relevant topics (Gauss's law) via a midterm exam by one of the authors (A. M.) who was not affiliated with any of the courses. All students had also completed the study of Calculus I and most of them were taking Calculus II at the time. They also completed the first semester of introductory physics, which includes a fair amount of mathematics. The interviewer ensured that students had not solved the problem prior to the interview (or at least did not recognize the problem when presented with it), so during the interviews students solved the problem for the first time while thinking out loud. The interviews were audio recorded, and during the interviews, A. M. took notes about key points in the interview to listen to carefully later. Students' work was also collected, and later, based on the notes, certain key points in the interview were transcribed. Two researchers looked at the important transcribed parts of the interviews and discussed the difficulties. There was some disagreement in the beginning in the interpretation of the data, but through discussions, any disagreements were resolved. We should point out that the research presented here used mixed methods and the main purpose of the think-aloud interviews was to help the researchers interpret the quantitative data.

### B. Results

**RQ1: Without any scaffolding support, to what extent are students able to make connections between the mathematical and graphical representations of the electric field in this problem?**—We found that only 26% of students plotted the electric field consistent with the mathematical expressions they found in all regions. Many students plotted the electric field consistently in some regions and inconsistently in others. It appeared from the plots that many students did not recognize that their expressions in different regions were showing a piecewise defined electric field. Therefore, we carried out think-aloud interviews to better understand the common difficulties students have in plotting the electric field for this problem.

**RQ2: What common difficulties do students have in transforming the electric field in this problem from the mathematical to the graphical representation and how common are these difficulties?**—Eight students who had completed the study of electrostatics were interviewed one-on-one using a think-aloud protocol. The interviews suggested two common difficulties students have in plotting the electric field in this problem: (1) not contemplating the behavior of the electric field in each region separately when plotting and globally plotting the electric field in all regions at once and (2) contemplating the behavior of the electric field in each region separately but plotting the "expected" behavior based upon qualitative reasoning rather than the mathematical expressions found.

We describe each difficulty in more detail and provide examples below.

**Student difficulty 1: Not contemplating the behavior of the electric field in each region separately when plotting and globally plotting the electric field in all regions at once.**—This difficulty was observed in interviews in which students insisted on plotting a continuous electric field without carefully contemplating what the behavior of the field is in each region or noticing the discontinuity in the electric field at the boundaries between regions. For example, one interviewed student, Alex, found the following expressions for the electric field:

$$r < a: \ E = kQ/r^2,$$
$$a < r < b: \ E = -2kQ/r^2,$$
$$b < r < c: \ E = -kQ/r^2,$$
$$r > c: \ E = -kQ/r^2.$$

Then, he mainly focused on the sign of the expression for the electric field in each region and the fact that the field must be continuous throughout to make his plot. He did not

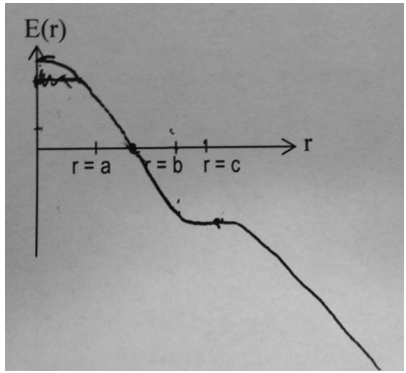

FIG. 4.   Electric field plotted by Alex (an interviewed student).

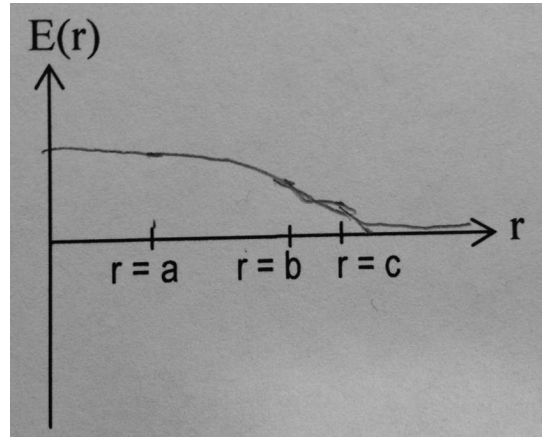

FIG. 5.   Electric field plotted by Charles (an interviewed student).

carefully check the behavior of the electric field, especially at the boundary between two regions where the field is discontinuous,

After observing his expressions in different regions, Alex started plotting the electric field (shown in Fig. 4). As he was plotting the electric field, he said,

> *E(r) is gonna start out really positive while it's in a [he means for r < a], it's gonna be more positive out here [points towards r close to 0], so it's gonna go like this [starts drawing a decreasing graph towards r = a] then start going down faster, hit that midpoint [points in between a and b where it is zero according to his plot], hit this point where r equals b, start leveling off until r equals c, and then drop down to negative infinity out here [region r > c].*

It was evident from the above quote that Alex was mainly focusing on the sign of the electric field and making sure that the field is continuous everywhere. He did not recognize that each of the different expressions has to be plotted separately in the appropriate region.

Another student, Charles, used a similar approach. He found the following expressions for the electric field:

$$r < a: \ E = kQ/r^2,$$
$$a < r < b: \ E = kQ/(b - a)^2,$$
$$b < r < c: \ E = kQ/(c - b)^2,$$
$$r > c: \ E = 0.$$

He then reasoned that at $r = a$, the electric field is larger than at $r = b$, which is larger than at $r = c$. He concluded that the field should continuously decrease starting from $r = 0$ and drew the graph shown in Fig. 5.

Similar to Alex, Charles did not focus on each region one by one while plotting and did not contemplate the behavior of the electric field in each region separately. He did not realize that there were discontinuities in the electric field at the boundaries between different regions. Instead, he plotted a continuous electric field by considering only several points in different regions ($r = a$, $r = b$, and $r = c$) at the same time.

He did, however, plot a zero electric field for $r > c$, which is consistent with his expression in that region.

Another interviewed student, Yara, used a similar approach when plotting the electric field. She found the following expressions:

$$r < a: \ E = kQr/a^3,$$
$$a < r < b: \ E = kQ/(r - a)^2,$$
$$b < r < c: \ E = 0,$$
$$r > c: \ E = -kQ/(r - c)^2.$$

Her plot is shown in Fig. 6.

Before plotting, she determined whether the electric field is negative or positive in each region and used the convention (which she made up) that if the electric field points away from the center, it is negative, and if it points towards the center, it is positive. She did this by reasoning conceptually and determined that for $r < a$ and $r > c$, the electric field points towards the center (positive according to her convention), and for $a < r < b$ and $b < r < c$, it points away from the center (negative according to her convention). As she was doing this, she plotted a few points (some of which can be seen in Fig. 6) to indicate that the electric field is negative or positive. She then plotted the electric field in all of the different regions together by only making sure that it has the expected sign (negative or positive) without thinking about what the exact behavior of the electric field is in each region (e.g., decreasing as $1/r$, constant, zero, etc.) or whether it is continuous at the boundaries between the regions.

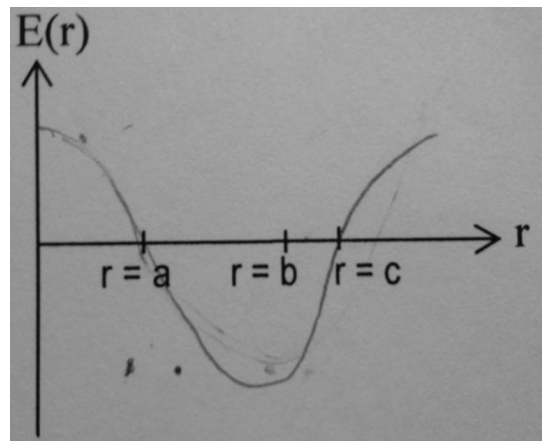

FIG. 6.   Electric field plotted by Yara (an interviewed student).

We note that both Charles and Yara were high-achieving students who performed well in the first semester calculus-based introductory physics course (both received an A−), and in the exam, which covered electrostatics in the second semester calculus-based physics courses, their scores were above average; Yara's score was actually above 90%. However, they still had great difficulty recognizing that the electric field should be plotted by considering its behavior in each region separately. When plotting the electric field, they did so globally (i.e., in all regions at once) without considering the behavior of the electric field in each region. These interviews suggested that in order to help students plot the electric field consistently, they should first be guided to recognize that they need to consider each region separately and identify what the behavior of the electric field is in each region and plot it accordingly.

**Student difficulty 2: Contemplating the behavior of the electric field in each region separately but plotting the "expected" behavior based upon qualitative reasoning rather than the mathematical expressions found.—** Some students recognized that they should consider the behavior of the electric field in each region separately while plotting, but they still drew plots that were inconsistent with their mathematical expression for the electric field. The reason for the inconsistency for these students was that they did not apply the same approach that was previously used for finding the expression of the electric field to make the corresponding plot in the same region (or vice versa if they drew the plot prior to writing down the mathematical expressions). Instead, they may adopt a qualitative approach while plotting (e.g., by recalling an expected behavior) but use a mathematical approach (e.g., by applying Gauss's law) when writing down an expression for the electric field. Because of the common introductory student difficulty in applying the mathematical form of Gauss's law correctly, more often than not, the qualitative reasoning of these students while plotting did not agree with their mathematical expression in the same region. For example, Sarah, in region $b < r < c$, started by reasoning qualitatively (it is possible that she recalled a memorized result):

> In there it should be zero because it's within a conductor." Then, after a short pause, she added "Now, if only I could find an expression for that.

It appeared that something like "$E = 0$" did not fit Sarah's conception of "expression," because she proceeded to try to find an expression with variables (or constants from the problem, $a$, $b$, $c$, $Q$) in it. She used Gauss's law mathematically, did so incorrectly, and obtained $E = -4\pi c^2 + 4\pi b^2$. At this point in the problem-solving process, Sarah did not recall (as she explicitly noted when reading the problem and when looking at the diagram provided) that $b$ and $c$ are the inner and outer radii of the spherical shell and that therefore they are not equal, which implies that her expression, $E = -4\pi c^2 + 4\pi b^2$, is not consistent with her qualitative expectation ($E = 0$). She was more inclined to trust an expression that followed from a mathematical procedure. However, when plotting the electric field, she plotted $E = 0$ (the behavior she was expecting from her qualitative reasoning).

It appears that some students like Sarah may have epistemological beliefs [52,73] that quantitative reasoning should be trusted when writing a mathematical expression and qualitative reasoning should be trusted when plotting. Sarah employed a similar approach in region $a < r < b$, in which, using Gauss's law mathematically, she obtained $E = -4\pi b^2 + 4\pi a^2$. But when plotting the electric field, she said,

> For r between distances a and b [...] we dropped off with E being proportional to $1/r^2$.

She then plotted a function that decreases in this way instead of plotting the expression she found through mathematically applying Gauss's law (a constant negative function).

Joe, another interviewed student, found a nonzero mathematical expression in region $b < r < c$, namely, $k|Q||\rho|/r^2$ (in this expression, $\rho$ refers to volume charge density, which Joe had not defined). However, when he plotted the field in this region, he said that the electric field should vanish because the negative charge, $-Q$, will be on the inner surface of the spherical shell (i.e., at $r = b$). Although he seemed to be aware that the field should vanish in this region while plotting, when writing an expression, he trusted the mathematical expression ($\sim 1/r^2$) he found.

Another interviewed student, James, in region $a < r < b$ included contributions from both the inner sphere and the outer spherical shell to obtain $2kQ/r^2$, which is the expression he wrote down for the electric field in that region. However, he noted that the situation given in the problem was a spherical capacitor and argued that the electric field should be constant:

> As we get farther away from [the edge of the sphere], [...] the outer circle's [outer spherical shell] field would get stronger in a way that the [net] field would remain constant anywhere between the two points [ $r = a$ and $r = b$].

He then plotted a constant, positive electric field between $r = a$ and $r = b$ (what he expected from his qualitative reasoning) instead of plotting the function he wrote down for the electric field in this region ($\sim 1/r^2$).

We note that the main difference between the two difficulties is that students with difficulty 1 do not contemplate the behavior of the electric field in each individual region separately, and plot the electric field in all regions at once, whereas students with difficulty 2 explicitly consider the behavior of the electric field in each region, but plot what they expect from qualitative reasoning instead of their expression. For example, Charles expected the field to decrease from one region to the next, but when plotting the field, he did it in all regions at once and plotted a decreasing function (Fig. 5) and did not stop to consider how that decrease should occur in each region (e.g., linear, proportional to $1/r$ or $1/r^2$). In contrast, Sarah explicitly considered how the electric field should behave in each region, and, after finding an expression, she plotted the field in each region separately instead of plotting it in all regions at once. However, instead of plotting her expression, she plotted the behavior she expected.

**How common are these two types of difficulties?**—In order to estimate how common the difficulties are, we analyzed the quantitative data to identify student responses that are likely to be a result of difficulty 1 or 2. Although it is difficult to precisely identify the number of students who do not contemplate the behavior of the electric field in each region separately based only on written work, the researchers obtained an estimate by counting the number of students who plotted continuous electric fields even though their expressions indicated that a discontinuity should be present. This is reasonable because during the interviews, the manner in which the students with difficulty 1 plotted the electric field indicated that they felt it should be continuous (some students explicitly noted this). We found that 51% of students who participated in the in-class study were likely to have this type of difficulty.

To estimate how many students plotted what their qualitative reasoning indicates the electric field should behave like instead of the mathematical expression they found, the researchers focused on all the cases in which there was an inconsistency between the mathematical and graphical representation and tried to identify whether it was possible that the inconsistency was due to difficulty 2. For example, in region $r < a$, a student may have found a nonzero expression for the electric field, but instead plotted a vanishing electric field. In this case, researchers included the student in the group that was likely having difficulty 2. Another example includes a student who found a nonlinear electric field in the same region, but plotted a linearly increasing electric field that is zero at the origin. In this case too. it was considered by the researchers that the inconsistency was likely a result of difficulty 2 because students may have incorrectly thought about the case in which the sphere is an insulating volume distribution of charge and not a conducting sphere. Also, for the region $a < r < b$, if a student plotted a constant (nonzero) electric field but found a different expression, it was also considered by researchers that this was likely due to difficulty 2 because students may have recognized, e.g., as James did, that the situation corresponds to a spherical capacitor, but incorrectly generalized from their knowledge of a parallel plate capacitor that this implies that the electric field is constant inside the capacitor. Our analysis of the written data indicates that including all such cases, in only 9% of the cases the inconsistency could be due to difficulty 2. It is important to keep in mind that the 9% included cases in which students plotted incorrect behaviors that they may have expected from qualitative reasoning. In only 4% of the cases, the plot a student drew was correct despite the fact that their expression was incorrect.

Written data in study 1 found that few students ($\sim 25\%$) were able to transform consistently from the mathematical to a graphical representation, and think-aloud interviews identified two difficulties that appeared to be common. Analysis of the written data (from the 65 student quizzes) suggested that difficulty 1 is a lot more common than difficulty 2 (51% compared to 9%). Therefore, in study 2, we designed two levels of scaffolding support to help reduce difficulty 1. The decision to focus on difficulty 1 was also partly influenced by the fact that the types of scaffolding required to help students with difficulty 2 are not necessarily the same as those required to reduce difficulty 1.

## III. STUDY 2

### A. Methodology and research questions

In study 2, we designed scaffolding support to help students recognize that the electric field in this problem has different expressions in different regions which must be plotted separately in each region. In order to design scaffolding supports, we first performed a cognitive task analysis [74,75] from an expert perspective. Cognitive task analysis is a technique designed to "yield information about the knowledge, thought processes, and goal structures that underlie observable task performance" [74], which, in the context of physics problem solving consists of identifying all of the individual thought processes required for students to be able to solve a problem. Instruction can help students develop those thought processes on their own by helping them learn to follow effective problem-solving techniques that experts employ [76]—an approach that has been shown to be effective in improving problem-solving performance [75,76]. Therefore, the intent of the scaffolding supports was to help students follow the steps an expert would when plotting the electric field for this Gauss's law problem. In addition, the design of the scaffolding supports was also influenced by our knowledge of how students who have difficulty 1 are reasoning while plotting the electric field. SL2 built on SL1 in that it included all the support of SL1 (plus additional support). We refer to the version of the problem used in study 1 as "scaffolding level 0" (SL0) because it did not involve any scaffolding.

**Scaffolding level 1 (SL1) design.**—Cognitive task analysis indicates that, when plotting the electric field for this problem situation, an expert would recognize that each expression for the electric field should be plotted in the appropriate region. Students with difficulty 1 plotted the electric field globally in all regions at once, which suggests than in order to help reduce this difficulty, scaffolding support should guide students to focus on one region at a time. This kind of support can have two benefits: (1) help students recognize that the behavior of the electric field should be considered in each region separately, and (2) reduce the cognitive load [56] associated with considering four regions at the same time.

Therefore, the first level of scaffolding (SL1) asked students to plot the electric field in each region separately before plotting it in part (b) for the problem situation shown in Fig. 1. The instructions provided to them in part (a) of the question were as follows:

(a) Write an expression for the electric field in each region and plot the electric field in that region on the coordinate axes shown (in the shaded regions, please do not draw).

For each region ($r < a$, $a < r < b$, etc.), right after calculating the expression for the electric field in that region, students were given coordinate axes with the irrelevant regions shaded out, as shown in Table I. This scaffolding level was designed to help students contemplate the electric field behavior in each region separately instead of in all regions at one time and, therefore, help reduce difficulty 1.

**Scaffolding level 2 (SL2) design.**—In addition to recognizing that each expression for the electric field should be plotted in the appropriate region, an expert is also likely to contemplate the end points of each expression in each region because this type of analysis provides explicit information about each expression (which is useful for plotting it). The plotting approaches of students with difficulty 1 suggested that they assumed that the electric field in this problem is continuous (some students explicitly said this). Therefore, providing scaffolding that helps students consider end points of different expressions in different regions can be beneficial, primarily in helping students recognize that the electric field is not continuous and plot it accordingly.

Students who received the second level of scaffolding (SL2) were provided all the support of SL1, and, in addition, they were asked to evaluate the electric field at the beginning, midpoint, and end point of each region immediately before plotting it in that region (see Table I). In addition to helping students recognize a discontinuity in the electric field, another potential benefit of SL2 is to help students make a consistent plot. For example, if a student is unsure about how a $1/r$ expression should be plotted, he or she could plug in values for several different $r$ to determine what the graph looks like. A student could potentially recognize how the plot should look like after calculating the function explicitly at three points, beginning, midpoint, and end point of the respective region.

We note that interviews with students who were thinking aloud while solving the SL1 and SL2 versions of the problem indicated that they were not confused by the instructions in the two scaffolded versions of the problem (i.e., it was clear to students where to plot the electric field, what the additional instructions meant, etc.). In addition, physics graduate students solved the two scaffolded versions of this problem and commented on the scaffolding provided (i.e., to what extent they expected it to be useful for introductory students). Some physics faculty members who had taught introductory physics recently were also shown the interventions and were asked to predict the effectiveness of the interventions. Both graduate students and faculty predicted that the interventions will help students connect their expressions with their plots better than students in the comparison group, with the majority predicting that students in the SL2 group will perform better in this regard than students in the SL1 group. Sometimes they specifically mentioned that the additional instructions would certainly be helpful for introductory students.

The scaffolded versions of the problem were given to 95 students in a traditionally taught calculus-based introductory physics course who were enrolled in three different recitation sections. The three recitation sections formed the comparison group and two scaffolding intervention groups for this investigation. All recitation sections were taught in a traditional manner by the same teaching assistant who worked out problems similar to the homework problems and then gave students a 20-minute quiz at the end of recitation. Students in all recitations attended the same lectures, were assigned the same homework, and had the same exams and quizzes.

Below, we delineate the research questions developed for the purposes of investigating the impact of different levels of scaffolding support on students' representational consistency and the connection between representational consistency and performance on this problem.

**RQ1: What is the impact of the medium level of scaffolding support on students' representational consistency on this problem?**—This research question was investigated by comparing the percentages of students who plotted the mathematical expressions they found consistently in the SL1 group with the SL0 group.

**RQ2: What are some mechanisms which may account for the impact of the medium level of scaffolding on students' representational consistency on this problem?**—In order to shed light on the possible mechanisms for how students are affected by the medium level of scaffolding on this problem, interviews were conducted

TABLE I.   Descriptions of the three scaffolding levels.

| Scaffolding level 0 (SL0) | Scaffolding level 1 (SL1) | Scaffolding level 2 (SL2) |
| --- | --- | --- |
| (a)(i) $r < a$<br>Write an expression for the electric field. | (a)(i) $r < a$<br>Write an expression for the electric field. | (a)(i) $r < a$<br>Write an expression for the electric field.<br>When $r = 0$, $E(r = 0) =$ _________<br>When $r = a/2$, $E(r = a/2) =$ _________<br>When $r = a$, $E(r \to a) =$ _________ |
| | Plot the electric field on coordinate axes provided (irrelevant regions shaded out).<br>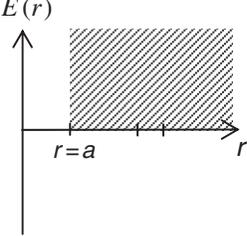 | Plot the electric field on coordinate axes provided (irrelevant regions shaded out).<br>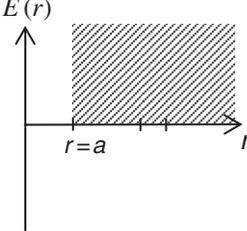 |
| Similar for other three regions, in parts (a)(ii), (a)(iii), and (a)(iv). | Similar for other three regions, in parts (a)(ii), (a)(iii), and (a)(iv). | Similar for other three regions, in parts (a)(ii), (a)(iii), and (a)(iv). |
| (b) Plot the electric field for all regions in (a):<br>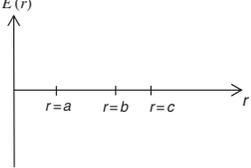 | 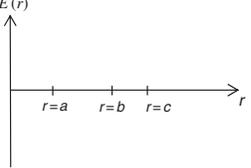 | 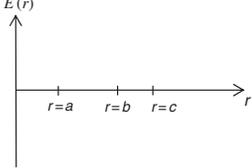 |

using a think-aloud protocol [72]. As mentioned earlier, eight interviews were conducted as part of study 1 with students who solved the problem with no scaffolding. Eight interviews were conducted with students who solved the SL1 version of the problem and their approaches to plotting the electric field were compared to the students provided with no scaffolding.

**RQ3: How does the additional scaffolding (SL2) compare to the medium level of scaffolding (SL1) in terms of students' representational consistency on this problem?**—This research question was investigated by comparing the percentages of students who plotted the mathematical expressions they found consistently in the SL2 group with the SL1 group. Since these students had to plot the electric field twice, once in each individual region separately (with irrelevant regions shaded out) and once in all regions combined, we did both comparisons (see Tables V and VI). (Note that it may seem unnecessary to perform the second comparison, but we should note that while an expert is likely to put together the plots he or she

drew in each region when plotting the electric field in all regions, students often do not do this. They sometimes have a final plot that is different from putting together the individual plots they have in each region.)

**RQ4: What mechanisms may be useful in explaining the impact of the additional scaffolding in SL2 compared to SL1 on introductory students' representational consistency on this problem?**—This question was investigated by conducting think-aloud interviews with seven students who solved the SL2 version of the problem and comparing their approaches to those of students who solved the SL1 of the problem while thinking aloud.

**RQ5: Do students who exhibit representational consistency perform better on this problem than students who do not?**—Previous research has found that students' representational consistency correlates with learning gains in mechanics [77] and that students who display superior skills in representing a physical problem improved problem-solving performance [78]. Motivated by these findings, we investigated whether students who were

consistent between the mathematical and graphical representations of the electric field exhibited improved problem-solving performance compared to students who were not consistent. In other words, we wanted to know whether students who were consistent were more likely to find the correct expressions for the electric field than students who were not.

A rubric was developed jointly by two researchers to quantify students' problem-solving performance. The two researchers agreed on the final version of the rubric prior to analyzing the data for this study and independently graded a randomly selected subset of 20% of the data which showed excellent agreement (interrater reliability better than 90%). The rubric used is summarized in Table II.

Table II shows that region $a < r < b$ was assigned 3 times as many points as each of the other regions. This consideration was made because region $a < r < b$ was the only one with a nonzero electric field. In finding the expression for the electric field in parts (a)(i) through (a)(iv), students were given 80% for the correct expression and 20% for the correct reasoning that led to that expression. For example, if a student wrote $E = k(Q/r^2)$ for the expression without any explanation in region $a < r < b$, he or she would obtain 24 or 30 points. If, instead, a student found the incorrect electric field and showed his or her work (which was also incorrect), he or she was awarded 10% of the credit for the attempt. In addition, to ensure that our grading scheme did not influence our results significantly, grading was also performed by weighing each of the regions in the same manner, i.e., assigning 10 points for finding the electric field in each of the parts (a)(i) through (a)(iv). All of the results we report for this study are the same using both methods, and we chose to describe the results using the grading method outlined above (which weighs region $a < r < b$ 3 times more than each other region).

It is important to keep in mind that students in SL1 and SL2 groups did not obtain any additional points for plotting the electric field in each region first, or for evaluating the electric field at the beginning, midpoint, and end point of each region. All the points given in the regions (a)(i) through (a)(iv), which required students to find the expression for the electric field, were based solely on the expressions students found.

## B. Results and discussion

Before analyzing the data that we used to answer the research questions delineated earlier, we analyzed all students' scores on the final exam and found that students in the three intervention groups exhibited similar performance with no statistically significant differences between different recitation sections (which were randomly assigned to be the comparison group and the two intervention groups in this study). We should also note that in the subsequent pages we compare different groups of students in terms of either (1) average scores or (2) percentages of students who are consistent. In order to compare average scores of two groups, we used $t$-tests, and in order to compare percentages chi-square tests were used [79], two commonly used approaches in physics education research. (Note that throughout this paper differences are considered significant if the $p$ value for comparing the groups is $\leq 0.05$.)

**RQ1: What is the impact of medium level of scaffolding support on students' representational consistency on this problem?**—We found that the medium level of scaffolding impacted students' representational consistency positively. In particular, we found that, compared to the SL0 group (comparison group), more students in the SL1 group were always consistent (i.e., consistent in all regions). Table III shows the percentage of students who were always consistent in the SL0 and SL1 groups. A chi-squared test [79] on these data shows that students in the SL1 group performed better in this respect than students in the SL0 group ($p = 0.020$). We note that the performance of students in SL0 in terms of representational consistency was nearly identical to the performance of students in study 1 who were also not provided with any scaffolding (29% compared to 26%). The SL1 intervention effectively doubled the percentage of students who plotted the electric field consistently from roughly 30% to roughly 60%.

**RQ2: What are some mechanisms which may account for the impact of the medium level of scaffolding on students' representational consistency on this problem?**—The interviews conducted in study 1 suggested that in order to help students plot the expressions they find consistently, scaffolding support may help students realize that the electric field has different behaviors in different regions which should each be plotted accordingly. The first level of scaffolding was designed to do just this: students were asked to plot the expression for the electric field they found in each region in order to provide a hint that each expression should be plotted separately. In addition, the irrelevant regions were shaded out to help focus students' attention on the appropriate

TABLE II. Summary of the scores assigned to each part of the problem.

| (a) Find an expression for the electric field | | | |
| --- | --- | --- | --- |
| (i) $r < a$ | (ii) $a < r < b$ | (iii) $b < r < c$ | (iv) $r > c$ |
| 10 points | 30 points | 10 points | 10 points |

TABLE III. Percentages (and numbers) of students from the SL0 and SL1 groups who were consistent in *all* regions, and the $p$ value for comparison between the groups via a chi-squared test.

| | Consistent | Not consistent |
| --- | --- | --- |
| SL0 | 29% (9) | 71% (22) |
| SL1 | 59% (16) | 41% (11) |
| $p$ value | 0.020 | |

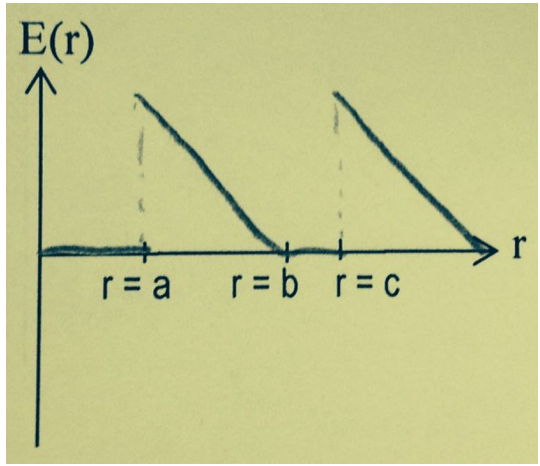

FIG. 7. Electric field plotted by Mike (an interviewed student).

regions. Interviews with students provided with this level of scaffolding suggested that it was indeed effective at helping them realize the piecewise nature of the electric field. When plotting the electric field, students were clearly contemplating its behavior in each region, and they would often say things like "in here [a particular region] it should drop off like $1/r^2$" or "in here, it is constant.". For example, Mike, an interviewed student, decided to find all of the expressions first before plotting them. He found the following expressions:

$$r < a: \ E = 0,$$
$$a < r < b: \ E = kQ/r,$$
$$b < r < c: \ E = 0,$$
$$r > c: \ E = kQ/r.$$

After finding them in each region he went back to the coordinate axes provided in each region and plotted each expression in those regions. Figure 7 shows the final graph he drew, which he obtained by putting together the four plots he drew in each region.

As he was plotting the electric field in each region, he clearly contemplated the behavior of the functions because he said,

*The graph for the first region] will just be a horizontal line at zero, and then the graph for (ii), the way I have it set up it should be... ok, so radius increases, that would decrease this [expression for electric field], so it should be a downward line […] and again, for (iii) it's gonna be a horizontal [line] and it's zero and for radius greater than c, the electric field should also decrease as the radius increases.*

Mike did not seem to know what a function proportional to $1/r$ looks like, and plotted it incorrectly. However, he did contemplate the behavior of the electric field in each region, and plotted each expression separately, which is what the SL1 intervention was designed to guide students to do. This approach (i.e., to plot each expression separately in each region), common among students who solved the SL1 version of the problem, is in stark contrast to some of the observed plotting approaches of students provided with no support who plotted the electric field globally in all regions at once by assuming it must be continuous (see, for example, Alex's approach from study 1). It is possible that the students in the comparison group had more information to process due to having to focus on more than one expression at a time, which increased their cognitive load [56] or distracted students and resulted in decreased performance. Interviews with students provided with the medium level of support suggested that, similar to Mike, they seemed to focus their attention only on the relevant region in which they were asked to plot the electric field, which helped them realize that they need to plot each expression separately. Thus, this scaffolding may have reduced the level of distraction and the possible cognitive load.

**RQ3: How does the additional scaffolding (SL2) compare to the medium level of scaffolding (SL1) in terms of students' representational consistency on this problem?**—Our research indicates that the higher level of scaffolding (SL2) actually had a detrimental effect (opposite of that intended) compared to the medium level of scaffolding (SL1). In particular, fewer students in the SL2 compared to the SL1 group were able to plot the electric field consistent with the expressions they found.

Students in the two scaffolding interventions were asked to plot the electric field in each region immediately after finding it and, in addition, they were provided with coordinate axes with the irrelevant regions shaded out. Table IV shows, in each of the four parts, the percentages of students who were consistent between the mathematical expressions they found and their plots, and the comparison via chi-square tests between students in SL1 and SL2 groups. The results show that students in the SL2 group were significantly less consistent than students in the SL1 group in all but the last part.

The data shown in Table IV suggest that sometimes students were consistent in one or more parts, but not all.

TABLE IV. Percentages (and numbers) of students from the SL1 and SL2 groups who were consistent between their plots and the expressions they found in each of the first four parts and $p$ values for comparing the two groups.

|  | (a)(i) | (a)(ii) | (a)(iii) | (a)(iv) |
|---|---|---|---|---|
| SL1 | 86% (24) | 67% (18) | 77% (20) | 69% (18) |
| SL2 | 55% (17) | 37% (11) | 32% (6) | 58% (11) |
| $p$ value | 0.012[a] | 0.024 | 0.005[a] | 0.534[a] |

[a]Because of one or more expected cell frequencies being less than 10, Fisher's exact test was performed instead of the standard chi-square test.

TABLE V. Percentages (and numbers) of students from the SL1 and SL2 groups who were consistent in *all* parts and *p* value for the comparison between the two groups via a chi-square test.

|  | Consistent | Not consistent |
|---|---|---|
| SL1 | 59% (16) | 41% (11) |
| SL2 | 24% (7) | 76% (22) |
| *p* value | 0.008 | |

We also investigated how many students were always consistent between the expressions they found and the plots they drew (i.e., consistent in all parts). This result is shown in Table V, which reveals that students in the SL2 group were significantly less consistent than students in the SL1 group. In fact, it appears that the additional instructions of SL2 compared to SL1 erased the benefit students derived from SL1 because a comparison between the data in Table V and Table III indicates that on this problem the percentage of students who plotted the electric field consistent with their mathematical expressions in all regions in the SL2 group is similar to the no scaffolding group (SL0).

**RQ4: What mechanisms may be useful in explaining the impact of the additional scaffolding of SL2 compared to SL1 on introductory students' representational consistency on this problem?**—The data discussed thus far indicated that students in the SL2 group exhibited less representational consistency between mathematical and graphical representations of the electric field than students in the SL1 group, which was surprising because the additional scaffolding of SL2 compared to SL1 was intended to help students be more consistent not less. As noted earlier, in order to obtain an in-depth account of how students were impacted by the two scaffolding interventions, we conducted think-aloud interviews with introductory students, seven of whom solved version SL2 of the problem and eight of whom solved version SL1 of the problem.

The interviews suggested that one possible cognitive mechanism that could partially explain this unexpected finding relates to a framework involving working memory [80,81]. In this framework, problems are solved by processing relevant information in the working memory. However, working memory has finite capacity (5–9 "slots") for any person, regardless of their intellectual capabilities [82]. In order to solve a problem, one has to determine the relevant information that must be processed at a given time in order to move forward with a solution. Some of the relevant information required for solving a problem must be retrieved from long-term memory (for example, relevant principles, such as Gauss's law or physics of conductors, mathematical information related to plotting functions, etc.), while other relevant information must be accessed via sensory buffers (eyes, ears, etc.). Experts generally solve problems by focusing on important features of the problem and by retrieving the appropriate information from their long-term memory [83,84], which has a well-organized knowledge hierarchy in their domain of expertise, and also by processing relevant information from the environment (e.g., from the problem statement and part of their solution written down so far). Novices, on the other hand, typically do not have a robust knowledge structure and their knowledge chunks are smaller [85]. Therefore, their working memory processing capacity is effectively reduced, which can lead to cognitive overload [56,85,86] and deteriorated performance. Alternatively, they may have enough cognitive resources (working memory slots) to process the relevant information while problem solving, but because they do not have a robust knowledge structure, they can focus on unimportant features of the problem [87,88]. They can get distracted by information that is not useful and lose track of relevant information which, if processed appropriately, can be helpful in solving the problem. This mechanism too can lead to deteriorated performance. We note, however, that it is difficult to distinguish between these cognitive mechanisms for deteriorated performance, because even in the case of cognitive overload, novices can lose track of relevant information since they may not have enough cognitive resources to process all that is required at a given time in order to move forward with a solution.

The additional instructions of SL2 as compared to SL1 as shown in Table I ask students to "find the limit of the electric field as $r$ approaches $a$, $b$, etc." in various regions and will henceforth be referred to as "limits." A cognitive task analysis from an expert perspective suggests that calculating these limits before plotting a function would be useful because they provide explicit information about the function at three distinct points, which is helpful for plotting it. To ensure that physics experts generally agree, we asked graduate students in a teaching assistant training class and several physics faculty members whether they expect that the additional instructions of SL2 would be helpful for introductory students when plotting the electric field (the graduate students solved the SL1 and SL2 versions of the problem before being asked this question). Some of the graduate students and physics faculty members noted that the additional instructions would definitely be helpful for introductory physics students. However, the interviews suggested that the introductory physics students in the SL2 group for whom the additional instructions were intended did not, in general, discern the relevance of these instructions to plotting the function in the next part. To them, evaluating the function at various points in a given interval appeared to be just another task that required additional information processing (which could have distracted them from processing relevant information or overloaded their working memory). Every single introductory physics student interviewed who had to evaluate the electric field at three points in each interval before plotting it lost track of some important information at one point or another: some students forgot to plot the electric field in a particular region, and some students went straight to evaluating the limits even before finding an expression for the electric field in that region. An example of losing track of important information comes from an interview with John. In finding the limits of the function in regions $r < a$ and $a < r < b$, John did not substitute the corresponding values for $r$. For example, he wrote $E(r \to a) = kQ/r^2$ without substituting $r = a$ into the expression. But then when he got to the first limit in region $b < r < c$ [$E(r \to b)$], after writing down an initial expression in which he did not substitute $r = b$, he suddenly realized on his own that he should substitute $r = b$:

> John: "Oh, should I plug in […] 'cause it's r approaching b?"
> Researcher: "I can't tell you that. […] What do you think?"
> John (after a short pause): "I'll just write it to be safe."

He then went back and changed all the previous limits where he had not substituted the corresponding values for $r$. Discussions with John suggested that the piece of information "when you find a limit of a function, you should substitute the value of the variable in that function" was present in his long-term memory, but he did not retrieve it until a particular point. For a while, he appeared to be focusing on and processing other information in the problem that was not helpful for figuring out the limits correctly. It is possible that part of the reason why John went back to his previous answers for the limits and changed them is that he was solving the problem while thinking aloud in an interview session, and in a quiz situation he would have moved on without stopping to question the correctness of his work.

Every single student interviewed who had to evaluate the expression at various points immediately before plotting overlooked something in a somewhat similar manner to John while solving the different parts of the problem, and the intended scaffolding involving explicit evaluation of the function at the three points was not discerned by the students as helpful for plotting the function. The additional instructions were just another chore for the students, which distracted them from processing the relevant information for plotting the electric field. During the interviews, students who solved the SL2 version of the problem were more likely than those who solved the SL1 version of the problem to go around in circles, repeat details mentioned earlier, and not retrieve relevant information from their memory at the appropriate time, even though that information was later revealed to be present in their long-term memory. It is also possible that because the additional instructions required more information processing, students' working memory got overloaded, which could lead to students struggling to process the relevant information and struggling to plot the mathematical expressions they found consistently.

**RQ5: Do students who exhibit representational consistency on this problem perform better than students who do not?**—We found that representational consistency was associated with improved student performance in determining the correct expressions for the electric field in each region. Table VI shows the averages and standard deviations in each of the four parts, (a)(i) through (a)(iv), of students who were consistent in all parts ("consistent" in Table VI) and students who were not consistent in one or more parts ("not consistent" in Table VI), regardless of what intervention group they were in. Table VI also shows the $p$ values (obtained via $t$-tests) and effect sizes for comparing the performance of these groups of students (students who were consistent in all parts and students who were not consistent in one or more parts), which reveal that the students who were consistent outperformed the other students in every part. The $p$ values for the comparison between these groups are also very small and the effect sizes (Cohen's $d$) show large effects, three of them being above 1.0 (for the last three parts). We also conducted this analysis with the data in study 1 and found the same results. (Note that Cohen's $d$ is defined as the difference in means of the two groups one compares divided by the standard deviation of the population from which the samples were taken. In practice, the standard deviation of the population is almost never known and is most commonly estimated by the standard deviation of the control or comparison group.

TABLE VI. Numbers of students ($N$), averages (Avg.), and standard deviations (St. dev.) in each part in which the scores were based on expressions for students who were consistent in all parts ("Consistent") and for students who were not consistent in one or more parts ("Not consistent"), and $p$ values and effect sizes for comparison between these two groups of students.

|  | Part (a)(i) | | | Part (a)(ii) | | | Part (a)(iii) | | | Part (a)(iv) | | |
|---|---|---|---|---|---|---|---|---|---|---|---|---|
|  | $N$ | Avg. | St. dev. | $N$ | Avg. | St. dev. | $N$ | Avg. | St. dev. | $N$ | Avg. | St. dev. |
| Consistent | 32 | 6.3 | 4.0 | 32 | 7.1 | 3.6 | 32 | 5.8 | 4.2 | 32 | 6.4 | 4.6 |
| Not consistent | 55 | 3.5 | 3.9 | 55 | 2.9 | 3.3 | 55 | 1.5 | 3.0 | 55 | 2.0 | 3.7 |
| $p$ value | | 0.002 | | | <0.001 | | | <0.001 | | | <0.001 | |
| Effect size | | 0.73 | | | 1.24 | | | 1.20 | | | 1.05 | |

For the two treatment groups we compare here, one can estimate the population standard deviation by using a pooled standard deviation based on the two standard deviations of the samples being compared. This pooled standard deviation is defined as $\sigma_{\text{pooled}} = \sqrt{(\sigma_1^2 + \sigma_2^2)/2}$ and was used as an estimation of the population standard deviation [79].)

## IV. SUMMARY

We found that calculus-based introductory physics students have great difficulty translating from a mathematical to a graphical representation—only about one-quarter of the students without support plotted electric fields that were consistent with their expressions in all regions for the problem investigated here. Via think-aloud interviews, we also identified possible reasons for the poor student performance in translating the electric field from the mathematical to the corresponding graphical representation. Their difficulties were twofold: (1) students often did not contemplate the behavior of the electric field in each region separately and plotted a continuous function by only thinking about some vague global characteristics of how the electric field changes in different regions, e.g., the electric field is larger in one region than in another, the electric field is positive in some regions, negative in others etc. (this difficulty was most common among students provided with no support), and (2) students realized that the electric field has different behaviors in different regions; however, they plotted what their qualitative analysis indicated the electric field should be in a region instead of plotting the mathematical functions they found. Our written data in study 1 indicated that the first difficulty was significantly more prevalent than the second one.

Based on these findings, we developed two scaffolding support levels with the intention of reducing the number of students who have difficulty 1, or in other words, to help students recognize that the electric field is a piecewise defined function that must separately be plotted in each region. We found that providing additional scaffolding (by asking students to evaluate the electric field at the beginning, midpoint, and end point of each interval), although intended to help students be more consistent in plotting the electric field, had an adverse effect on their representational consistency for the problem in electrostatics with spherical symmetry discussed here that can be solved using Gauss's law. Physics graduate students and physics faculty members (experts for this study) remarked that they expected the additional scaffolding to be helpful. However, think-aloud interviews conducted with introductory physics students suggested that they did not discern the relevance of these additional instructions in the SL2 version and often got distracted by this additional task that they treated as a chore (or they may have had cognitive overload due to the need to attend to additional instructions provided while engaged in solving the problem). An important instructional implication of this finding is that, although cognitive task analysis from an expert perspective can be valuable and provide insight into the scaffolding support that might be beneficial in a given situation, it is important to assess how students perceive the scaffolding support designed, for example, by observing how students solve the problem with scaffolding support in think-aloud interviews. As we found in our investigation, a high level of scaffolding support from an expert perspective may not always benefit students and it can even lead to deteriorated performance.

We also found that asking introductory students to plot the electric field in each region immediately after finding an expression for it in that region (the SL1 intervention) impacted students positively, resulting in improved performance in translating between mathematical and graphical representations for the Gauss's law problem analyzed. Interviews suggested that the improved representational consistency was partly due to students recognizing that the electric field behaves differently in different regions and plotting it accordingly.

## ACKNOWLEDGMENTS

We would like to thank the National Science Foundation for Award No. DUE-1524575. Also, we are extremely grateful to Professor F. Reif, Professor J. Levy, and Professor R. P. Devaty, and all the members of the Physics Education Research group at University of Pittsburgh for very helpful discussions and/or feedback on the manuscript.

---